# Spin- and Isospin-Dependent Momentum Distributions in Fermi Liquids at Non-zero Temperatures

M.Serhan [1] and M.L.Ristig [2]

[1] Department of Physics, Al al-Bayt University, Al-Mafraq, Jordan

[2] Institut für theoretische Physik, Universität zu Köln, D-50937 Köln, Germany

## Abstract

We explore the structure of momentum distributions of Fermi liquids such as completely polarized $^3$He, unpolarized liquid $^3$He, and nuclear matter at nonzero temperatures. The study employs correlated density matrix theory and adapts the algorithm to deal with spin- and isospin-dependent correlations. The analysis is based on the factor decomposition of the one-body reduced density matrix. The decomposition permits to distinguish between statistical correlations and dynamic (direct) correlations. Together with the concept of renormalized fermions the formal results open the pathway to investigate the thermal boundaries of normal Fermi phases within correlated density matrix theory. We also discuss possible transitions from normal phases to anomalous fermion phases triggered by statistical correlations or by periodic phase-phase structures.

**Keywords**: Momentum distribution; Fermi liquids; density-matrix theory; spin-dependent correlations.

## 1. Introduction

Strongly correlated Fermi systems are of great interest. They exist in normal Fermi liquid phases but may also develop anomalous phases with multi connected Fermi surfaces or phases of fermion condensates [1]. It needs sophisticated microscopic many-body theories to explore such phases or to study the transition region from the normal liquid phase to the anomalous quantum states. Such topological phase transitions may be heralded by the development of anomalous or divergent features of the existing strong correlations in the normal phases close to their phase boundaries. It is therefore of interest to study in detail the properties of one-body and two-body reduced density matrix elements of strongly correlated fermion systems.

    In this study we concentrate on the analysis of the structure of the one-body density matrix elements and their spin (isospin) dependence as functions of particle density $\rho$ and temperature $T$. The study is performed in close analogy to Refs [2-4] that employed Correlated Density Matrix (CDM) theory.

    A study of one-body reduced elements is also worthwhile since their Fourier transforms represent the single particle momentum distribution in the fermion medium. It plays a role central to the understanding of systems of quantum particles. The distributions are of interest for most of the subjects of research in modern physics: systems of atoms, interacting electron systems, quantum liquids, etc. [5].

Early theoretical studies of momentum distributions have been based on Ursell-Mayer cluster expansions [6-8] to explore the correlated ground states of Bose and Fermi systems. Ref [9] reports on recent applications of this technique to liquid $^3$He. Ref [10] provides an early review on theoretical momentum distributions of various Bose and Fermi systems at zero temperature within cluster approaches and hypernetted-chain (HNC) techniques.

This study employs systematically CDM theory at nonzero temperatures (including the limit $T \to 0$) [2, 11,12] and generalizes the formalism to include the influence of spin/ isospin dependent correlations on the momentum distributions.

Section 2 gives a brief summary of the basic CDM results for the momentum distribution $n(k)$ and on the one-body reduced density matrix elements $n(r)$ for spatially correlated spinless fermions. This formalism is then generalized to include spin (isospin) effects (Sec. 3). In Sec. 4 the algorithm is brought into detailed explicit form by projecting the expressions onto coordinate space. Exploiting the particular structure of the one-body density $n(12)$ we evaluate the spin-isospin dependent statistical factor $N_0(12)$ and the phase-phase direct correlation function $Q(12)$ in Section 5. The concluding Sec. 6 presents a brief summary of the formal results and discusses possible transitions from normal phases to anomalous phases within CDM theory.

## 2. Completely polarized liquid $^3$He

In this Section we briefly summarize basic results of CDM theory on the correlated one-body reduced density matrix of spinless fermions at thermal equilibrium[2]. Based on these results we then develop their generalization to deal with spin (isospin) correlation effects in symmetric liquid $^3$He and nuclear matter at nonzero temperatures. In completely polarized liquid $^3$He all $N$ spins are parallel, the single spin level is not degenerate and the occupation is characterized by $\nu = 1$. All correlations depend only on the radial distance $r$ in coordinate space. In this case the one-body reduced density matrix at given density and temperature is represented by the CDM result [2],

$$n(r) = n_c N_0(r) \exp[-Q(r)] , \qquad (1)$$

with the strength factor $n_c = \exp Q(r = 0)$. Function $N_0(r)$ represents the statistical distribution effected by the correlated fermion system and the phase-phase correlation function $Q(r)$ describes the dynamic effects directly generated by the fermion-fermion interaction. The dimensionless Fourier transform of the elements (1) is the momentum distribution of a single fermion in the many-fermion medium,

$$n(k) = \rho \int n(r) e^{i\vec{k}\cdot\vec{r}} dr . \qquad (2)$$

The sum rule

$$\sum_k n(k) = N \qquad (3)$$

expresses the conservation of the total number of fermions in the liquid. Relation (3) yields the consequence $n(r) = 1$ at $r = 0$ and leads to the unit-normalization $N_0(r = 0) = 1$ for the statistical distribution factor appearing in result (1).

The CDM algorithm [2] provides the analytic tools to evaluate functions $N_0(r)$ and $Q(r)$ and their Fourier transforms $N_0(k)$ and $Q(k)$, respectively. They are given by decompositions in terms of so-called ($QQ$) contributions of nodal ($N$) and elementary ($E$) types,

$$N_0(k) = \Gamma_{cc}(k) - N_{QQcc}(k) - E_{QQcc}(k), \tag{4}$$

and, respectively,

$$-Q(k) = N_{QQdd}(k) + E_{QQdd}(k). \tag{5}$$

The subscripts indicate the exchange or circular ($cc$) nature of the statistical function $N_0(k)$ generated by the statistical input factor $\Gamma_{cc}(k)$. In Eq. (5) the subscripts ($dd$) characterize the dynamic ($dd$) origin of the phase-phase structure function $Q(k)$.

The nodal portions are products in momentum space of generalized structure functions $S_{Qcc}(k)$, $S_{Qdd}(k)$, $S_{Qde}(k)$, and their corresponding non-nodal components $X_{Qcc}(k)$, $X_{Qdd}(k)$ and $X_{Qde}(k)$ appearing in the decompositions

$$S_{Qdd}(k) = X_{Qdd}(k) + N_{Qdd}(k)$$
$$S_{Qde}(k) = X_{Qde}(k) + N_{Qde}(k), \tag{6}$$

$$S_{Qcc}(k) = X_{Qcc}(k) + N_{Qcc}(k) - \Gamma_{Qcc}(k). \tag{7}$$

The product representation of functions $N_{QQcc}(k)$ and $N_{QQdd}(k)$ read

$$N_{QQcc}(k) = [X_{Qcc}(k) - \Gamma_{cc}(k)]S_{Qcc}(k) \tag{8}$$

and

$$N_{QQdd}(k) = [X_{Qdd}(k) + X_{Qde}(k)]S_{Qdd}(k) + X_{Qdd}(k)S_{Qde}(k). \tag{9}$$

A set of $HNC_Q$ equations permits to calculate these quantities by employing HNC techniques [2].

## 3. Spin (Isospin) dependent formalism

To generalize the CDM formalism for dealing with equilibrium properties of unpolarized liquid $^3$He (degeneracy $\nu = 2$) and symmetric nuclear matter (spin/isospin degeneracy $\nu = 4$) we may proceed along the lines reported in Ref [4] for the liquid structure function $S(k)$. Here, we generalize functions $N_0(r)$ and $Q(r)$ of Sec. 2 to properly include the effects of spin (isospin) dependence. To do this the statistical component and the dynamic function are assumed to depend also on spin (isospin) variables, $Q(r, \sigma_1, \sigma_2)$, $N_0(r, \sigma_1, \sigma_2)$ or, respectively, $Q(r, \sigma_1, \sigma_2; \tau_1, \tau_2)$ and $N_0(r, \sigma_1, \sigma_2; \tau_1, \tau_2)$. The variables are the relative distance $r$, and the spin isospin $\sigma$ and/ or $\tau$, taken, conventionally, in z-direction and ignoring the x and y components of the spin and isospin vectors in coordinate space [4]. For convenience we may abbreviate this dependence by writing $N_0(12)$, $Q(12)$, and any other function $L(r_{12}, \sigma_1, \sigma_2; \tau_1, \tau_2) \equiv L(12)$.

The generalization of Eq. (1) then reads in spin (isospin) space and coordinate space

$$n(12) = n_c N_0(12) \exp[-Q(12)]. \tag{10}$$

The statistical distribution function $N_0(12)$ and the direct-direct phase function $Q(12)$ are given by the decompositions in analogy to relations (4) and (5)

$$N_0(12) = \Gamma_{cc}(12) - N_{QQcc}(12) - E_{QQcc}(12) \tag{11}$$

and

$$-Q(12) = N_{QQdd}(12) + E_{QQdd}(12), \tag{12}$$

respectively. As in the case $\nu = 1$ the nodal components $N_{QQcc}$ and $N_{QQdd}$ can be evaluated from the generalized modified distribution functions $G_{Qdd}(12)$, $G_{Qde}(12)$, and $G_{Qcc}(12)$. They can be constructed by applying $HNC_Q$ techniques in analogy to the HNC procedure for functions $G_{dd}(12)$, $G_{de}(12)$, and $G_{cc}(12)$ performed in Ref. [4]. Decomposing the $(Q_{dd})$, $(Q_{de})$, and $(Q_{cc})$ correlation functions into nodal and non-nodal components,

$$\begin{aligned} G_{Qdd}(12) &= X_{Qdd}(12) + N_{Qdd}(12), \\ G_{Qde}(12) &= X_{Qde}(12) + N_{Qde}(12), \\ G_{Qcc}(12) &= X_{Qcc}(12) + N_{Qcc}(12) - \Gamma_{cc}(12) \end{aligned} \tag{13}$$

we are then able to write down the modified $HNC_Q$ equations for unpolarized liquid $^3$He and symmetric nuclear matter ($\nu = 2$ and $\nu = 4$, respectively),

$$\begin{aligned} 1 + G_{Qdd}(12) &= \exp[\tfrac{1}{2}u(12) + N_{Qdd}(12) + E_{Qdd}(12)], \\ G_{Qde}(12) &= [1 + G_{Qdd}(12)][N_{Qde}(12) + E_{Qde}(12)] - N_{Qde}(12), \\ G_{Qcc}(12) &= [1 + G_{Qdd}(12)][N_{Qcc}(12) + E_{Qcc}(12) - \Gamma_{cc}(12)]. \end{aligned} \tag{14}$$

omitting the subscript $Q$ and replacing the dynamic generator $u_Q(12) \equiv \frac{1}{2}u(12)$ by the pseudopotential $u(12)$ the set (14) specializes to the HNC equations of Ref. [4] for the radial distribution function $g(12) = 1 + G(12)$.

The temperature dependence of the functions appearing in the set (14) is induced by the ansätze for the exchange input quantity $\Gamma_{cc}(12)$ and for the dynamic input potential $u(12)$ as appropriate functions of $T$.

The chain equations complementing the hypernet equations (14) for case $\nu = 4$ read

$$N_{Qdd}(12) = \frac{\rho}{\nu} \sum_{\sigma_3,\tau_3} \int dr_3 \{[X_{Qdd}(13) + X_{Qde}(13)]G_{dd}(23) + X_{Qdd}(13)G_{de}(23)\},$$

$$N_{Qde}(12) = \frac{\rho}{\nu} \sum_{\sigma_3\tau_3} \int dr_3 \{[X_{Qdd}(13) + X_{Qde}(13)]G_{de}(23) + X_{Qdd}(13)G_{ee}(23)\}, \quad (15)$$

$$N_{Qcc}(12) = \rho \sum_{\sigma_3\tau_3} \int dr_3 [X_{Qcc}(13) - \Gamma_{cc}(13)]G_{cc}(23).$$

For unpolarized liquid $^3$He ($\nu = 2$) the summation over the isospin $\tau_3$ must be omitted. The input functions $G_{dd}(12)$, $G_{de}(12)$, $G_{ee}(12)$, and $G_{cc}(12)$ are the solutions of the corresponding HNC equations reported in Ref. [4].

## 4. Projection onto coordinate space

Th explicit expressions of the spin (isospin) components for parallel spins and antiparallel spins or for spin-independent and spin-dependent decompositions can be constructed employing the same classification procedure as performed in Ref. [4]. The decomposition of any function $L(12)$ follows the rule ($\nu = 2$)

$$L(12) = L^p(r)\delta_{\sigma_1\sigma_2} + L^a(r)[1 - \delta_{\sigma_1\sigma_2}] \quad (16)$$

where $L^p$ is the spin-parallel (superscript $p$) component in coordinate space and $L^a(r)$ is the anti-parallel portion. The spin-independent and spin-dependent projections of $L(12)$ onto coordinate space are

$$L(12) = L(r) + L^\sigma(r)\sigma_1\sigma_2 \quad (17)$$

with $\sigma = \pm 1$. The analogous decompositions hold in case $\nu = 4$ also for the isospin projections with $\tau = \pm 1$.

Working with the decomposition (16) for functions $G_{Qdd}(12)$, $G_{Qde}(12)$, $G_{Qcc}(12)$ enables us to decouple the associated hypernet equations (14) into two sets. The first set relates only the spin-parallel pieces with each other, i. e.,

$$1 + G_{Qdd}^p(r) = \exp[\tfrac{1}{2}u^p(r) + N_{Qdd}^p(r) + E_{Qdd}^p(r)],$$
$$G_{Qde}^p(r) = [1 + G_{Qdd}^p(r)][N_{Qde}^p(r) + E_{Qde}^p(r)] - N_{Qde}^p(r), \quad (18)$$

$$G_{Qcc}^p(r) = [1 + Q_{Qdd}^p(r)][N_{Qcc}^p(r) + E_{Qcc}^p(r)] - \Gamma_{cc}(r). \quad (19)$$

The second set for the spin-antiparallel components $G^a{}_{Qdd}(r)$ and $G^a{}_{Qde}(r)$ is given by Eqs. (18) where the superscript $p$ is to be replaced by the superscript $a$. The statistical function $\Gamma_{cc}(12)$ is of the special form

$$\Gamma_{cc}(12) = \Gamma_{cc}(r)\delta_{\sigma_1\sigma_2} \quad (20)$$

since the Pauli exclusion principle holds only for parallel (identical) spins. Thus $\Gamma_{cc}(r) \equiv \Gamma_{cc}^p(r)$ and $\Gamma_{cc}^a(r) \equiv 0$.

The chain equations (15), however, decouple into two sets by adopting the decomposition (17). This possibility is brought about by the relationship [4]

$$\frac{1}{\nu}\sum_{\sigma_3} L(13)K(32) = L(r_{13})K(r_{23}) + L^\sigma(r_{13})K^\sigma(r_{23})\sigma_1\sigma_2 \qquad (21)$$

($\nu = 2$) that is valid for any functions $L$ (12) and $K$ (12). An analogous relation holds in isospin space.

The decoupling of set (15) is best formulated in momentum space, i. e. by representing the r-space convolutions in Eqs. (15) by products in k-space. Thus, with the Fourier transform

$$L(k) = \rho \int dr L(r) e^{i\vec{k}\cdot\vec{r}} \qquad (22)$$

for any function $L(r)$ or $L^\sigma(r)$ we find the projected chain equations

$$\begin{aligned}N_{Qdd}(k) &= [X_{Qdd}(k) + X_{Qde}(k)]S_{dd}(k) + X_{Qdd}(k)S_{de}(k)\\ N_{Qde}(k) &= [X_{Qdd}(k) + X_{Qde}(k)]S_{de}(k) + X_{Qdd}(k)S_{ee}(k)\end{aligned} \qquad (23)$$

that hold true for the spin-independent components. Attaching the superscript $\sigma$ to the structure functions $S_{\alpha\beta}(k)$ and the non-nodal pieces $X_{Q\alpha\beta}(k)$ with $\alpha\beta = dd, de, ee$ Eqs. (23) then represent the corresponding chain equations for the spin-dependent (superscript $\sigma$) correlations. Due to the particular spin dependence of the statistical generator $\Gamma_{cc}$(12), Eq. (20), the circular exchange equation reads in k-space

$$N_{Qcc}(k) = [X_{Qcc}(k) - \Gamma_{cc}(k)]S_{cc}(k). \qquad (24)$$

Eq.(24) is the chain relation for the statistical ($cc$) structure function $S_{Qcc}(k)$. With the decomposition

$$S_{Qcc}(k) = X_{Qcc}(k) + N_{Qcc}(k) - \Gamma_{cc}(k) \qquad (25)$$

we may derive, from Eq. (24), the result

$$S_{Qcc}(k) = [X_{Qcc}(k) - \Gamma_{cc}(k)][1 + S_{cc}(k)]. \qquad (26)$$

## 5. Spin dependence of momentum distributions

The momentum distribution of a single fermion in a Fermi liquid with spin degeneracy ($\nu = 2$) is given by the Fourier transform

$$n(k) = \sum_{\sigma_1}\rho\int dr n(12)e^{i\vec{k}\cdot\vec{r}}. \qquad (27)$$

For the case $\nu = 4$ a summation must be also performed over the isospin, $\tau_1 = \pm 1$. To evaluate the distribution (27) we begin with the expression (10) and a detailed analysis of the spin-isospin dependent factor $N_0$ (12) and the phase-phase correlation function $Q$(12). These functions are given by the decompositions (11) and (12), respectively. The nodal portions $N_{QQcc}$(12) and $N_{QQdd}$(12) contained in Eqs. (11) and

(12) can be traced back to the modified structure functions $S_{Qcc}(k)$, $S_{Qdd}(k)$ and their non-nodal parts $X_{Qcc}$ and $X_{Qdd}$. They are the analogues of the nodal functions $N_{Qcc}(12)$ and $N_{Qdd}(12)$ analyzed in Sec. 3 with the chain results (15). The chain relations read now ($\nu = 2$)

$$N_{QQcc}(12) = \rho \sum_{\sigma_3} \int dr_3 [X_{Qcc}(13) - \Gamma_{cc}(13)] G_{Qcc}(23) \tag{28}$$

$$N_{QQdd}(12) = \frac{\rho}{\nu} \sum_{\sigma_3} \int dr_3 \{[X_{Qdd}(13) + X_{Qde}(13)] G_{Qdd}(23) + X_{Qdd}(13) G_{Qde}(23)\}. \tag{29}$$

The projection onto $r$-space can be performed as in Sec. 4 for $N_{Qcc}$ and $N_{Qdd}$ with the result (24) and (23), respectively. In the present case we arrive at the chain equation for $N_{QQcc}(12)$ in $k$-space,

$$N_{QQcc}(k) = [X_{Qcc}(k) - \Gamma_{cc}(k)] S_{Qcc}(k). \tag{30}$$

Result (30) can be equivalently cast in the form

$$N_{QQcc}(k) = S_{Qcc}^2(k)[1 + S_{cc}(k)]^{-1} \tag{31}$$

due to Eq. (26). The same procedure is adopted to arrive at the $k$-space representation of Eq. (29). We decompose function $N_{QQdd}(12)$ according to Eq. (17),

$$N_{QQdd}(12) = N_{QQdd}(r) + N^\sigma_{QQdd}(r)\sigma_1\sigma_2 \tag{32}$$

and write the two chain equations for the spin-independent $N_{QQdd}(r)$ component and the spin-dependent portion $N^\sigma_{QQdd}(r)$ in $k$-space by properly modifying the first equation of set (23),

$$\begin{aligned} N_{QQdd}(k) &= [X_{Qdd}(k) + X_{Qde}(k)] S_{Qdd}(k) + X_{Qdd}(k) S_{Qde}(k), \\ N^\sigma_{QQdd}(k) &= [X^\sigma_{Qdd}(k) + X^\sigma_{Qde}(k)] S^\sigma_{Qdd}(k) + X^\sigma_{Qdd}(k) S^\sigma_{Qde}(k). \end{aligned} \tag{33}$$

Exploiting the analytic results derived in Sections 3 and 4 we are thus able to perform numerical calculations on functions (33) with the HNC technique.

After the detailed analysis of the building blocks $N_0(12)$ and $Q(12)$ we may now explicitly express the one-body density matrix $n(12)$ in terms of spin-parallel components.

The exchange factor has nonzero elements for parallel spins and vanishes for antiparallel spins,

$$N_0(12) = N_0(r)\delta_{\sigma_1\sigma_2} \tag{34}$$

with

$$N_0(r) = \Gamma_{cc}(r) - N_{QQcc}(r) - E_{QQcc}(r). \tag{35}$$

For free Fermi gases the Fourier transform of Eq. (35) specializes to the exact result

$$N_0(k) = \Gamma_{cc}(k)[1+\Gamma_{cc}(k)]^{-1} \tag{36}$$

due to relation (31) where $S_{QQcc}(k) = S_{cc}(k)$ and $N_{QQcc}(k) = N_{cc}(k)$. It is evident in this case that expression (36) may be identified with the familiar Fermi distribution of free fermions. We therefore have

$$n(12) = n(r)\delta_{\sigma_1\sigma_2} = N_0(r)\delta_{\sigma_1\sigma_2}. \tag{37}$$

The momentum distribution (27) of non-interacting fermions is therefore the Fourier transform $n(k) = N_0(k)$. The Fermi energy $\varepsilon_0(k_F)$ follows from the sum rule

$$\frac{1}{N}\sum_k \nu n(k) = \nu n(r=0) = 1. \tag{38}$$

For interacting fermions we must account for the strong dynamic phase-phase correlation function $Q(12)$ directly generated by the fermion-fermion interaction. Quantity $Q(12)$ influences also (indirectly) the exchange function $N_0(r)$ via relation (35) and (30) or (31). The direct correlation effects of $Q(12)$ on the one-body reduced density matrix (10) are described by the exponential factor $\delta_{\sigma_1\sigma_2}\exp[-Q(12)] = \exp[-Q(12)\delta_{\sigma_1\sigma_2}]$, i. e., by the spin-aligned component $Q^p(r)$.

To be more explicit we decompose $Q(12) = Q(r) + Q^\sigma(r)\sigma_1\sigma_2$ with the spin-independent component

$$-Q(r) = N_{QQdd}(r) + E_{QQdd}(r) \tag{39}$$

and the spin-dependent part

$$-Q^\sigma(r) = N^\sigma_{QQdd}(r) + E^\sigma_{QQdd}(r). \tag{40}$$

where the Fourier transforms of the nodal terms are to be calculated via the chain relations (33). The spin-parallel correlation function is therefore given by the sum $Q(r) + Q^\sigma(r)$. The final result for the one-body reduced density matrix is therefore

$$n(12) = \delta_{\sigma_1\sigma_2} N_0(r) n_c \exp[-Q^p(r)] \tag{41}$$

with the strength factor $n_c = \exp Q^p(r=0)$. The momentum distribution of spin-dependent spatial correlations is then the Fourier transform (27).

## 6. Discussion

In this study we have analyzed the structure of the momentum distribution of a single fermion in a strongly correlated normal Fermi medium and the associated one-body reduced density matrix as functions of density, temperature, and spin-degeneracy ($\nu = 1, 2, 4$). The analysis has been performed by employing CDM theory in conjunction with HNC techniques. The results generalize earlier work within cluster expansions and work on correlated ground states ($T = 0$) by Fermi-HNC procedures. CDM theory has been adopted since it leads to a very fruitful separation of exchange (or statistical) effects from dynamic effects directly generated by the interparticle interactions. This separation is evident in the factor representation (10) with the statistical distribution $N_0(12)$ and the dynamic phase-phase cluster expansions, work on correlated ground states by (Fermi) HNC procedures and other correlation function $Q(12)$.

Further, as we have demonstrated in Refs [4, 12] CDM theory permits application of the concept of renormalized fermions for studying particle exchange properties independent of exponential phase correlations described by function $Q(12)$. These renormalized fermions form a homogeneous background system of free particles obeying familiar Fermi statistics but have effective single-particle energies $\varepsilon_{qp}(k)$ with effective masses depending on temperature and momentum $\hbar k$ [11, 12]

$$n_{qp}(k) = \Gamma_{qp}(k)[1+\Gamma_{qp}(k)]^{-1}, \tag{42}$$

$$\Gamma_{qp}(k) = \exp\beta[\mu_{qp} - \varepsilon_{qp}(k)]. \tag{43}$$

A close study of the distributions $n_{qp}(k)$, $N_0(k)$, and $Q(k)$ and their Fourier inverse may permit to explore the phase boundaries of the normal fermion fluid and to signal transitions to possible anomalous fermion phases. For example, one could imagine phase transitions that are triggered by statistical effects and other types of transitions caused by anomalous features of the dynamical phase-phase correlation function $Q(r)$.

In normal Fermi liquids function $Q(r)$ is short-ranged. However, depending on the particle-particle interaction the phase-phase correlations may form a periodic structure of spatial oscillations with a characteristic wavelength or wave vector $k_0$. Equivalently, we may say that the phase-phase structure function $Q(k)$ exhibits a condensate at a particular momentum $\hbar k_0$.

A different kind of anomalous phase may be created if the statistical function $N_0(r)$ or the momentum distribution (42) is characterized by a non-monotonous effective energy $\varepsilon_{qp}(k)$ of the normalized fermions. The excitation energy in expression (43) may have the shape of a Landau curve as is familiar from the phonon-roton excitations of liquid $^4$He with a maximum at wavenumber $k_M$ (maxon) and a minimum at $k_R$ (roton). In this scenario the chemical potential $\mu_{qp}$ of the statistical factor (43) may intersect the excitation curve $\varepsilon_{qp}(k)$ at three values of wave numbers $k_1 < k_2 < k_F$, with $k_1 < k_M < k_2$ and $k_2 < k_M < k_F$. In the first interval $(k_1, k_2)$ the effective energy $\varepsilon_{qp}(k)$ is larger than the effective chemical potential $\mu_{qp}$. In contrast, in the interval $(k_2, k_F)$ we find $\varepsilon_{qp}(k) < \mu_{qp}$. Consequently, at $T = 0$ the Fermi surface becomes biconnected, i. e., $n_{qp}(k) = 0$ for $k_1 < k < k_2$. This anomaly characterizes therefore a particular type of topological phase transition and state.

## Acknowledgment

(M. S.) acknowledges gratefully a DAAD scholarship during this work.